# Emerging Optics from Structured Nanoscale Optical Cavities

Danqing Wang[1]*, Ankun Yang[2]

[1]Miller Institute, University of California, Berkeley, Berkeley, California, 94720, USA

[2]Department of Mechanical Engineering, Oakland University, Rochester, Michigan 48309, USA

*Correspondence to: danqingwang2018@u.northwestern.edu

**ABSTRACT**

Miniaturized and rationally assembled nanostructures exhibit extraordinarily distinct physical properties beyond their individual units. This review will focus on structured small-scale optical cavities that show unique electromagnetic near fields and collective optical coupling. By harnessing different material systems and structural designs, various light-matter interactions can be engineered, such as nanoscale lasing, nonlinear optics, exciton-polariton coupling, and energy harvesting. Key device performance of nanoscale lasers, including low power threshold, optical multiplexing, and electrical pump, will be discussed. This review will also cover emerging applications of nanoscale optical cavities in quantum engineering and topological photonics. Structured nanocavities can serve as a scalable platform for integrated photonic circuits and hybrid quantum photonic systems.

Keywords:

optical cavities, nanolasers, plasmonics, surface lattice resonance, quantum photonics, light-matter interactions

# 1. INTRODUCTION

## 1.1 Overview

Small-scale photonic cavities offer advances in enhancing the density of optical states for extraordinary light-matter interactions. Traditional Fabry–Pérot cavities that provide optical feedback for lasing involve bulky elements of a pair of mirrors. Developments in nanofabrication techniques enable scalable patterning of a wide range of nanomaterial designs over macroscale (> $cm^2$) areas and fine control over the nanoscale dimensions (~10 nm). Rationally assembled nanostructures can exhibit synergistic physical and chemical properties not achievable in their individual units, which offers immense opportunities in various structural designs of small-scale optical cavities for optical interactions.

This review article focuses on miniaturized photonic cavities for integrated photonics and quantum photonics. First, we introduce the development of miniaturized optical cavities and the system of plasmonic nanoparticle (NP) lattices. Second, we describe the recent development of structural designs and material systems of small-scale optical cavities for solid-state laser devices. Third, we discuss the key device functionalities desired for small-scale lasers. Finally, this review highlights emerging applications of small cavities in quantum engineering and new cavity systems based on theoretical advances in topological states. Small-scale optical cavities offer exciting prospects for solid-state devices, integrated photonics, and quantum nanophotonics.

## 1.2 Miniaturized optical cavities

Laser is a ubiquitously used light source in checkout counters at retail stores, computer printing, and medical therapy. Conventional lasers rely on macroscale optical cavities such as a pair of mirrors that support the Fabry–Pérot modes. Over the past decades, there has been an on-demand need for miniaturized optical cavities [1]. One key figure-of-merit for quantifying the enhancement of an emitter within an optical cavity is Purcell factor $F = \frac{3}{4\pi^2}\left(\frac{Q}{V}\right)\left(\frac{\lambda}{2n}\right)^3$, where $V$ is the mode volume, $Q$ is the cavity quality factor, $\lambda$ is the wavelength, and $n$ is the refractive index [2, 3]. Compared to a pair of mirrors that sustain Fabry–Pérot modes, miniaturized cavities can support smaller cavity mode volume $V$ close to the diffraction limit, which offers the potential for stronger Purcell enhancement for various light-matter interactions.

**Figure 1** shows an overview of small-scale optical cavities that evolve from microscale, nanoscale to picoscale. Microscale optical cavities based on dielectric materials have been widely developed in the past decades as optical feedback for lasing that utilize photonic crystal lattices [4, 5], self-assembled nanowires [6], and microring resonators [7, 8]. An electrically driven photonic crystal laser was realized at room temperature with a low threshold current (~260 µA) (**Figure 1a**) [9]. However, the overall mode volume of dielectric cavities is diffraction limited and larger than or on the order of the wavelength of the emission light. Access to sub-wavelength light confinement is limited by optical cavity designs.

Plasmonic nanocavities that support surface plasmons at the interface of metal and dielectrics provide a strategy to break the diffraction limit and confine the light at sub-wavelength scales. Since the concept of the spaser (surface plasmon amplification by stimulated emission of radiation) [10] was proposed two decades ago, plasmon nanolasers have been developed based on different structural configurations [11-14]. Subwavelength lasing was first realized by semiconducting nanowires on a metal film at cryogenic temperatures using surface plasmon polariton (SPP) modes as optical feedback (**Figure 1b**) [15]. Room-temperature lasing was enabled by improved cavity mode quality in semiconducting nano-squares with total internal reflection [16]. Integration of atomic-smooth metal films from epitaxial growth process further decreased optical scattering losses and lowered the lasing thresholds [17, 18].

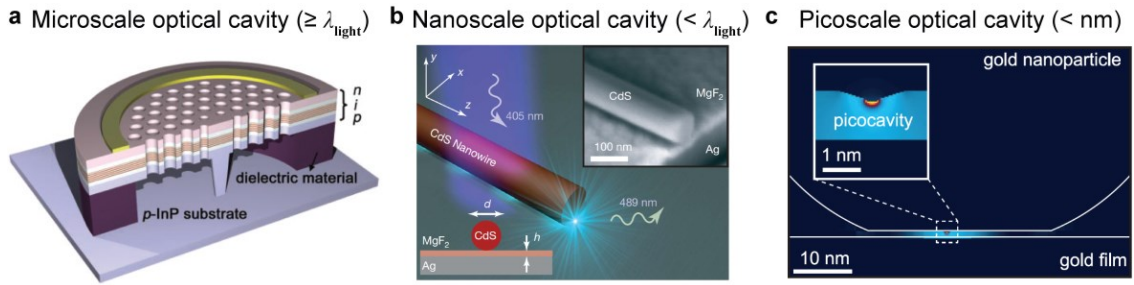

**Figure 1. Structural evolution and development of small-scale optical cavities.** (**a**) Microscale photonic crystal for electrically driven lasing. (**b**) Nanoscale laser based on nanowire-on-film plasmonic cavities. (**c**) Picocavity based on gold nanoparticles on a gold film. Panel (a) adapted from ref. 9 with permission; copyright 2004 AAAS. Panel (b) adapted from ref. 15 with permission; copyright 2009 Nature Publishing. Panel (c) adapted from ref. 21 with permission; copyright 2016 Nature Publishing.

Beyond nanoscale optical cavities, recent advances in plasmonic picocavities provide a route to further reduce the cavity mode volume [19]. Based on NP-on-mirror geometry as nanocavities, single-molecule strong coupling was accessed at room temperature and ambient conditions [20]. Individual atomic features inside such plasmonic gap can further localize light to sub-nm volumes and form a picocavity stabilized at cryogenic temperatures [21] (**Figure 1c**). The ultrasmall cavity mode volume contributes to significant enhancement of optomechanical coupling between the picocavity and molecular vibration states. Recently, Raman vibrational frequency shift induced by the optical spring effect was probed in picocavities by surface-enhanced Raman scattering [22]. Such extreme optical field confinement offers prospects in atom-cavity coupling and modulated photochemistry processes.

**1.3 Plasmonic nanocavity arrays**

Rationally assembled nanostructures exhibit distinct optical properties beyond their individual units. Periodic metal nanostructures show long-range diffractive interactions when the lattice spacing is close to the wavelength of incident light. Unlike a single metal NP that supports relatively broad localized surface plasmons (LSPs, full-width at half-maximum, FWHM > 50 nm), the collective optical interactions of metal NPs in a lattice induce sharp and intense surface lattice resonances (SLRs) [23-28]. SLRs originate from the coupling of LSPs of individual NPs to the Bragg modes defined in a lattice, which leads to reduced resonance linewidths (FWHM < 5 nm) with a characteristic Fano resonance line shape (**Figure 2a**). Experimentally, large-scale nanofabrication tools that combine multi-step processes of phase-shifting lithography, metal deposition, and Si etching enabled metal NP lattices over $cm^2$ areas with a programmable NP size and lattice spacing (**Figure 2b**). Measured optical band structure generated from angle-resolved transmission plots suggests that slow light can be trapped at SLR band-edge states at zero wavevectors with a nearly-zero group velocity ($v = \delta w/\delta k$) (**Figure 2c**).

Plasmonic NP lattices exhibit strongly enhanced optical fields within the sub-wavelength vicinity of the NP, which are two orders of magnitude higher than that of individual metal NPs. For plasmonic NP lattices, the combination of a subwavelength mode volume and high cavity mode quality from suppressed radiative loss contributes to a strong Purcell enhancement [29]. By integrating dye molecules as gain materials, lasing emission was demonstrated from the plasmonic nanocavity arrays at room temperature [30] (**Figure 2d**). In addition, plasmonic lattices provide a scalable platform to interface with different materials for various light-matter interactions, including enhanced nonlinear optics, exciton-polariton coupling, and photocatalysis processes [31-34].

Beyond single-lattice NP arrays, advances have been made in structural designs of nanoscale optical cavities for long-range optical interactions. Multi-modal lasing was realized with control over the different colors in a single device [35]. This system harnesses NP superlattices—finite arrays of metal NPs grouped into microscale arrays—integrated with liquid gain to produce multiple lasing colors (**Figure 2e**). Compared to traditional lasers, the multiscale superlattice cavities exhibit distinctive characteristics, including stable multi-modal lasing and well-controlled output. Varying superlattice geometries provides a robust way to manipulate the emission wavelengths, numbers of lasing beams, and angles at which the beam emits from the surface. Multi-modal lasing from superlattice arrays offers critical insights into the design of multi-modal lasing based on manipulating the physical geometry and optical band structures of NP lattices.

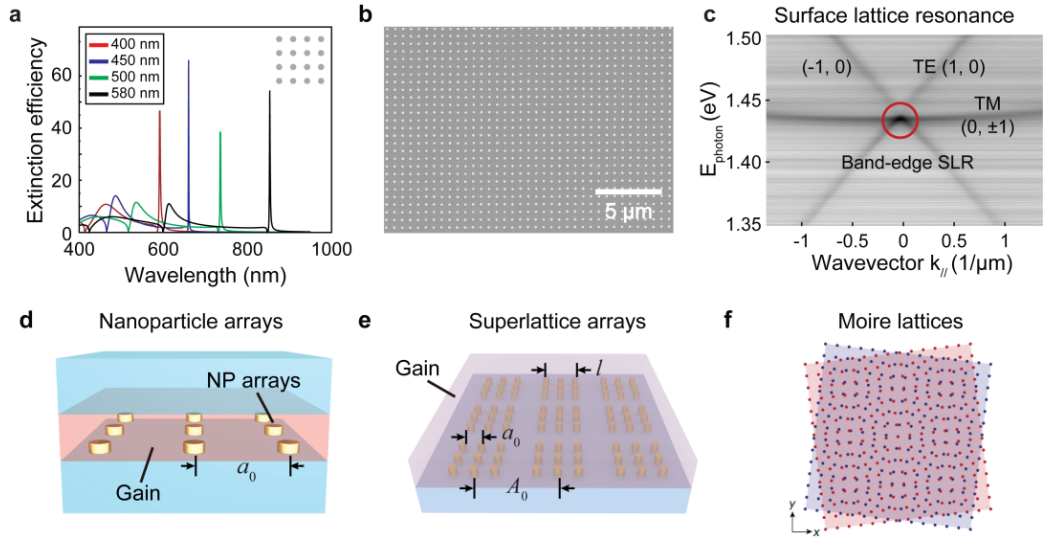

**Figure 2. Structured plasmonic nanocavity arrays based on metal nanoparticle lattices.** (**a**) Calculated extinction efficiency spectra of 2D Ag NP lattices with sharp SLRs based on coupled dipole method. (**b**) Scanning electron microscope (SEM) image of large-scale Au NP lattices. (**c**) Energy dispersion diagram of SLRs. (**d**) Scheme of single-lattice NP arrays integrated with gain medium for lasing. (**e**) Scheme of multiscale superlattice arrays for multimodal lasing. (**f**) Scheme of Moiré NP lattices for interlayer optical interactions. Panel (a) adapted from ref. 23 with permission; copyright 2004 American Institute of Physics. Panels (b-c) adapted from ref. 32 with permission; copyright 2019 American Chemical Society. Panel (d) adapted from ref. 30 with permission; copyright 2013 Nature Publishing. Panel (e) adapted from ref. 35 with permission; copyright 2017 Nature Publishing. Panel (f) adapted from ref. 36 with permission; copyright 2023 Nature Publishing.

Superposing two or more periodic structures to form moiré patterns emerges as a new platform to confine and manipulate light with a new degree of freedom in the vertical dimension. Recent advance in ultralong-range coupling between photonic lattices was made in bilayer and multilayer moiré structures mediated by dark SLRs (**Figure 2f**). 2D Au NP lattices enable twist-angle-controlled directional lasing emission for two lattices vertically separated by up to hundreds of microns [36]. Such far-field inter-lattice coupling between plasmonic lattices suggests a new out-of-plane dimension for optical modulation at small scales.

## 2. NEW MATERIALS SYSTEM

### 2.1 Plasmonic materials for nanocavities

The combination of NP size, lattice spacing, and materials component is critical for producing ultrasharp SLRs. Early demonstration of narrow SLRs relies on Au and Ag NPs, which sustain strong plasmon resonances and relatively low optical losses for device applications in the visible and near-infrared regimes. Because of their intrinsic material properties, NPs made of various metals exhibit different LSP resonances [37]. **Figures 3a-b** show the reported optical constants for different metals measured at 0.2 – 2 µm [38-41], where real part of the relative permittivity describes the strength of the polarization induced by an external electric field and imaginary part of the relative permittivity describes the losses encountered in polarizing the material. **Figure 3c** depicts the spectral ranges of metal NP resonances from UV to near-infrared. Here we highlight recent developments in plasmonic materials systems for accessing different SLR wavelength regimes and various light-matter interactions.

Compared to Au and Ag, Al exhibits lower optical losses (indicated by the low imaginary part of permittivity $\varepsilon$) in the UV to the visible regime. Al NP arrays integrated with an elastomeric slab exhibited high-quality SLRs (FWHM: 3–7 nm) continuously tailored over a large wavelength range (>100 nm) and at regime (500-600 nm) not possible for Au or Ag lattices [42]. Toggling between dipolar and quadrupolar SLRs is possible by stretching the hexagonal lattice along different symmetry directions. Both dipolar and hybrid quadrupolar SLRs from Al NPs can serve as optical feedback for nanolasing [43]. In addition, Al showed comparable lasing properties to Au, even at the interband transition wavelength regime.

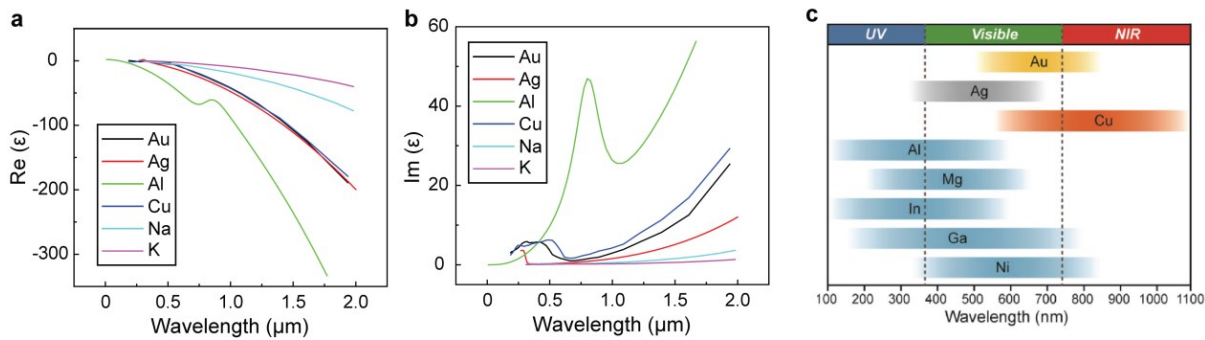

**Figure 3. Materials for plasmonic nanocavities.** (**a**) Real part of relative permittivity for Au, Ag, Al, Cu, Na and K. (**b**) Imaginary part of relative permittivity for Au, Ag, Al, Cu, Na and K. (**c**) The spectral ranges of plasmonic metal NPs. Panel (c) adapted from ref. 37 with permission; copyright 2018 Wiley-VCH. The optical constants of Au and Cu were taken from Johnson and Christy measurements [38], Al from Palik [39], Ag from Yang and Raschke [40], and Na and K from Smith [41].

Post-thermal annealing of metal NP lattices results in NPs with uniform shapes and smooth surfaces, which narrows the SLR linewidths toward the theoretical limit. Copper (Cu) is commonly used as a catalysis material for photochemistry processes. By using Cu NP lattices as catalytic substrates in a chemical vapor deposition process, the surface oxidation of Cu was prevented by conformally-grown graphene, and ultranarrow SLRs were realized with FWHM as small as 2 nm [44]. More systematic studies based on stable SLR linewidths showed that interfacial engineering at the metal-molecule interface is critical to tune the lasing thresholds [45]. Bimetallic core−shell (Cu−Pt) NP arrays can also modulate photoelectrocatalytic activity and enhance hydrogen evolution reactions by two folds, which offers exciting possibilities for nanocavity-enhanced photochemical and photoelectrochemical processes [46].

Alkali metals, particularly sodium (Na) and potassium (K), have long been considered ideal plasmonic materials [47]. Recent developments were made utilizing Na-based nanostructures as new types of plasmonic cavities. Based on the optical constants in Figures 3a-b, Na shows a better figure-of-merit ($-\varepsilon_1/\varepsilon_2$, where $\varepsilon_1$ and $\varepsilon_2$ are real and imaginary parts of the optical permittivity, respectively) compared to Ag in the near-infrared regime. A Na-based plasmonic nanolaser using a whispering-gallery mode was achieved at room temperature at 1257 nm [48]. The reported lasing threshold (140 kW/cm$^2$) was lower than other plasmonic nanolasers based on metal films in the near-infrared. A scalable fabrication method was developed for Na nanostructures by combining phase-shift photolithography and a thermo-assisted spin-coating process [49]. SPP resonance with linewidth as narrow as 9.3 nm was observed for the Na nanopit arrays, and the linewidth narrowing is sustained from the visible toward the NIR regime. In addition, K and NaK alloys have been explored to expand the material library for alkali metal plasmonics and liquid plasmonics [50]. In particular, Na-K alloy has a rich phase diagram with composition and temperature tunability, potentially enabling active metamaterials and reconfigurable photonic devices.

## 2.2 Active gain materials

**Figure 4** shows the library of gain materials used to couple to optical nanocavities for different light-matter interactions. Dye molecules integrated within a solid-state polyurethane matrix were exploited for the first observation of nanolasing from plasmonic nanocavity arrays [30] (**Figure 4a**). According to theoretical studies on the stimulated emission process based on a four-level semi-quantum model [51, 52], dye molecules close to the vicinity of the plasmonic spots contribute mostly to the lasing action. By further dissolving the dye molecules in a liquid organic solvent (such as dimethyl sulfoxide, ethylene glycol, and benzyl alcohol), the Brownian motion contributes to refreshing the dye molecules at the nanocavity hotspots for better laser device stability [53].

Unlike randomly-positioned dye molecules, spatially defined crystalline emitter materials of metal–organic frameworks (MOFs) conformally coated on metal NP lattices can support hybrid

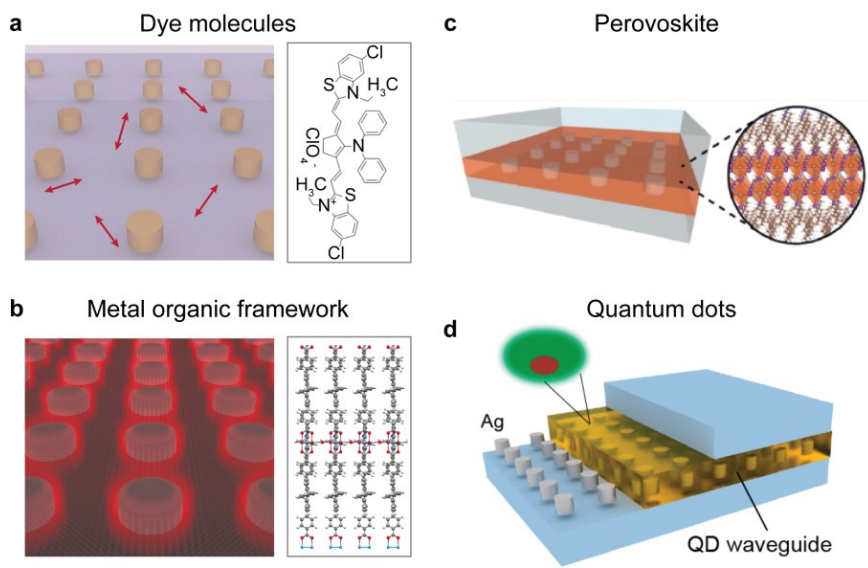

**Figure 4. Advances in gain material systems for coupling to nanophotonic cavities.** (**a**) Organic dye molecules. (**b**) Metal organic frameworks. (**c**) 2D layered perovskite. (**d**) Core-shell quantum dots. Panels (a-b) adapted from ref. 32 with permission; copyright 2019 American Chemical Society. Panel (c) adapted from ref. 57 with permission; copyright 2022 ACS Publications. Panel (d) adapted from ref. 60 with permission; copyright 2020 American Chemical Society.

exciton–plasmon modes. A 16-fold enhancement of photoluminescence intensity and a 2-fold decrease in exciton lifetime was observed for Zn-porphyrin MOFs coupled to Ag NP lattices [54] (**Figure 4b**). The mode splitting can be interpreted as modulation of the emission profile by the SLRs, which provides a strategy for spatially and temporally enhancing the properties of luminescent solid-state emitters. More distinct strong coupling was later achieved between Pd−P MOF (Zn paddlewheel node, Pd-metalated porphyrin ligand) and Ag NP lattices with characteristic upper polariton and lower polariton bands [55, 56]. Varying the SLR wavelength by changing the surrounding solvent across the $Q_1$ and $Q_2$ exciton bands of the MOF results in reversible evolution between weak and strong coupling regimes. Strong coupling with a signature avoided mode crossing in the dispersion diagram has also been observed from 2D

Ruddlesden−Popper perovskites thin film coupled to Al NP lattices [57] (**Figure 4c**). The coupling strength can be tuned by the film thickness and superstrate refractive index.

As solution-processable nanocrystals, colloidal quantum dots exhibit high photoluminescence quantum yields and long-term photostability [58]. The high-quality optical feedback supported at W-SLR sidebands leads to donut-shaped lasing patterns [59] (**Figure 4d**). Interestingly, either radial or azimuthal polarization can be controlled by the quantum dot film thickness, which determines the overlap of gain emission profile to SLRs hybridized with either transverse electric or transverse magnetic mode of the air/quantum-dot/silica waveguide. By exploiting high-symmetry points in plasmonic NP lattices (Δ and M point in the Brillouin zone), the direction of off-normal emission from quantum dot lasing was engineered to be as large as 19° [60]. By varying quantum dot concentrations, a weak-to-strong coupling transition was also observed in quantum dot thin films coupled to Ag NP lattices [61].

## 3. KEY DEVICE FUNCTIONALITIES

Compared to electronic circuits, photonic chips that use light as the medium offer a possible solution to faster communication and computing with their potential for larger bandwidth, faster modulation speed, and lower power consumption [62]. Si-based photonic chips are currently limited by device miniaturization, optical loss, and signal multiplexing [63]. In this section, we discuss the on-demand device functionalities of small-scale lasers as potential light sources for next-generation photonic circuits.

### 3.1 Low power threshold

Continuous-wave (CW) lasing at room temperature is critical for integration with optoelectronic devices and facile modulation of optical interactions. Insufficient gain concerning losses and thermal instabilities in nanocavities, however, have limited nanoscale lasers to pulsed pump sources and low-temperature operation. Stable, efficient gain materials under CW pump and high-quality optical nanocavities are critical for achieving low-threshold nanoscale lasing.

A materials platform was developed to realize continuous-wave upconverting lasing at room temperature with record-low thresholds and high photostability (**Figure 5a**). Selective, single-color lasing was observed based on $Yb^{3+}/Er^{3+}$-co-doped upconverting NPs conformally coated on Ag nanopillar arrays (**Figure 5b**). The intense electromagnetic fields localized in the vicinity of the nanopillars result in an ultralow lasing threshold [29]. Under CW pump, lasing emission occurred at the SLR wavelength defined by the NP lattice (**Figure 5c**). Above lasing threshold, a significant increase in rising slope ($s$ = 4.4) was observed in the input–output curve compared to spontaneous emission ($s$ = 2.2), along with simultaneous narrowing of the lasing emission linewidth (**Figure 5d**).

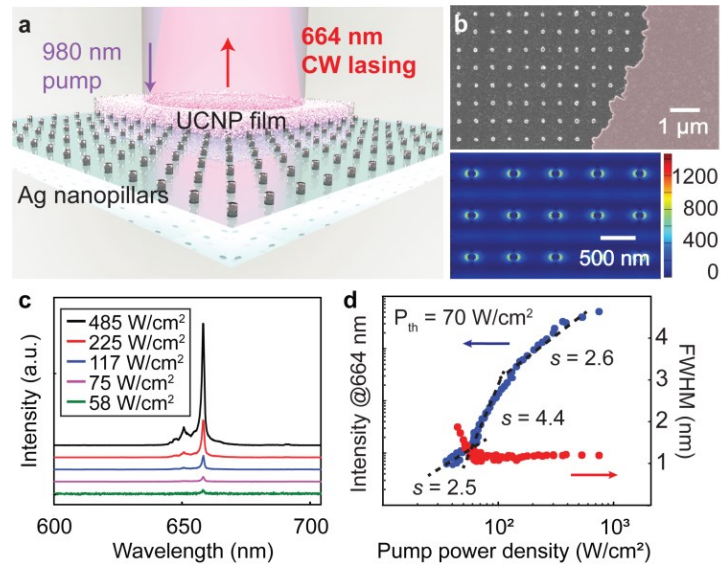

**Figure 5. Ultralow-threshold, continuous-wave upconverting nanolasing at room temperature.** (**a**) Schematic of the UCNP thin film coated on top of Ag arrays with lattice spacing $a_0$ = 450 nm. (**b**) SEM image showing the Ag nanopillar array with partial conformal coating with core-shell UCNPs ($NaYF_4:Yb^{3+}, Er^{3+}$). Bottom figure shows representative near-field $|E|^2$ plot for the 450-nm spaced Ag nanopillars. (**c**) Power-dependent lasing spectra from a SLR resonance at $\lambda$ = 664 nm. (**d**) Input–output curves in log–log scale showing emission linewidth narrowing and a lasing threshold $P_{th}$ of 70 W/cm². Figures adapted from ref. 29 with permission; copyright 2019 Nature Publishing.

**Table 1** shows a summary chart for different operation conditions and reported lasing thresholds for nanoscale lasers. The reported lasing threshold as small as 29 Wcm$^{-2}$ from upconverting nanolaser was >200-fold lower than that of upconverting microresonators [64] and orders of magnitude reduction over that of nanodisk-on-film [16, 65] or nanowire-on-film [18, 66] plasmon nanolasers at room temperature. Such a low lasing threshold comes from a synergistic materials system of high-quality nanocavities with suppressed radiative loss and photostable emitters with high quantum efficiency. Upconverting nanolasers provide a directional, ultra-stable output at visible wavelengths under near-infrared pumping, which offers prospects in medicine, quantum science, and next-generation photonic devices.

**Table 1**. **Summary of operation condition and power thresholds based on different cavity designs of nanoscale lasers**. Figure adapted from ref. 29 with permission; copyright 2019 Nature Publishing.

| Laser type | Year | Pump (pulse length, operation frequency) | Temperature | Lasing Threshold |
|---|---|---|---|---|
| Plasmon nanowire-on-film laser [15] | 2009 | 100 fs, 80 MHz | < 10 K | 50 MW/cm$^2$ |
| Plasmon microdisk-on-film laser [16] | 2011 | 100 fs, 10 kHz | Room temperature | 3000 MW/cm$^2$ |
| Plasmon nanowire-on-film laser [17] | 2012 | CW | Liquid nitrogen | 3.7 kW/cm$^2$ |
| Plasmon nanoparticle array laser [30] | 2013 | 100 fs, 1 kHz | Room temperature | 1 mJ/cm$^2$ |
| Plasmon nanohole array laser [67] | 2013 | CW | < 150 K | 5 kW/cm$^2$ |
| Plasmon nanowire-on-film laser [66] | 2014 | 150 fs, 800 kHz | Room temperature | 0.2 mJ/cm$^2$ |
| UV plasmon nanowire-on-film laser [18] | 2014 | 10 ns, 100 kHz | Room temperature | 1 MW/cm$^2$ |
| Plasmon nanodisk-on-film laser [65] | 2017 | 4.5 ns, 1 kHz | Room temperature | 10 kW/cm$^2$ |
| Photonic UCNP microbead laser [64] | 2018 | CW | Room temperature | 50 kW/cm$^2$ |
| Upconverting plasmon nanolaser [29] | 2019 | CW | Room temperature | 29 W/cm$^2$ |
| Plasmon nanodisk-on-film laser [48] | 2020 | 5 ns, 12 kHz | Room temperature | 140 kW/cm$^2$ |

**3.2 Output tunability and optical multiplexing**

Wavelength tunability of the lasing emission is a critical driver for practical device applications. The emission of conventional lasers is typically fixed at the time of device fabrication, and manipulating output color requires intricate optical designs. Tunable lasers that can emit at different wavelengths from a single device are essential for applications ranging from multiplexed signal processing to multi-color biomedical imaging.

Real-time tunable lasing emission can readily be achieved with plasmonic nanocavities because the SLR wavelengths are tunable by the surrounding refractive index ($n$) [53]. The linewidth of SLR modes is also tolerant to minor index mismatches between the superstrate and the substrate ($\Delta n < 0.05$) [68]. Tunable lasing was demonstrated by integrating a plasmonic NP lattice within a microfluidic channel (**Figure 6a**). By injecting dye solutions in different solvents that exhibit varying $n$, the lasing wavelength can be shifted over the entire emission bandwidth of the dye (**Figure 6b**). In contrast to a solid-state matrix, Brownian motion of liquid dye molecules results in a constant refreshing of active emitters in the plasmonic hot spots. Therefore, nanoscale lasers showed reduced photo-oxidation and photobleaching of dyes and a longer device lifetime [69, 70].

Inspired by the color change mechanism of chameleons, mechanical control of the lasing color was realized based on plasmonic NP lattices in a stretchable polymer matrix (**Figure 6c**). Liquid dye molecules dissolved in organic solvents were used as the gain media, ensuring close contact with the NPs upon stretching of the substrate [71]. Large metal NPs (260 nm in diameter) arranged in a lattice (600 nm in spacing) produce high-quality hybrid quadrupolar SLRs with narrow resonance linewidth (FWHM < 5 nm). When pumped with a pulsed laser, the single-color nanolaser emits at the near-infrared regime normal to the sample surface. The lasing peak shifts toward the longer wavelength side when the device is stretched, and exhibits excellent recovery after releasing the strain (**Figure 6d**). By stretching and relaxing the substrate, dynamic tuning of the lasing emission color was obtained. The mechanically tunable nanolaser could provide advances in future wearable and flexible optical displays for portable electronic devices.

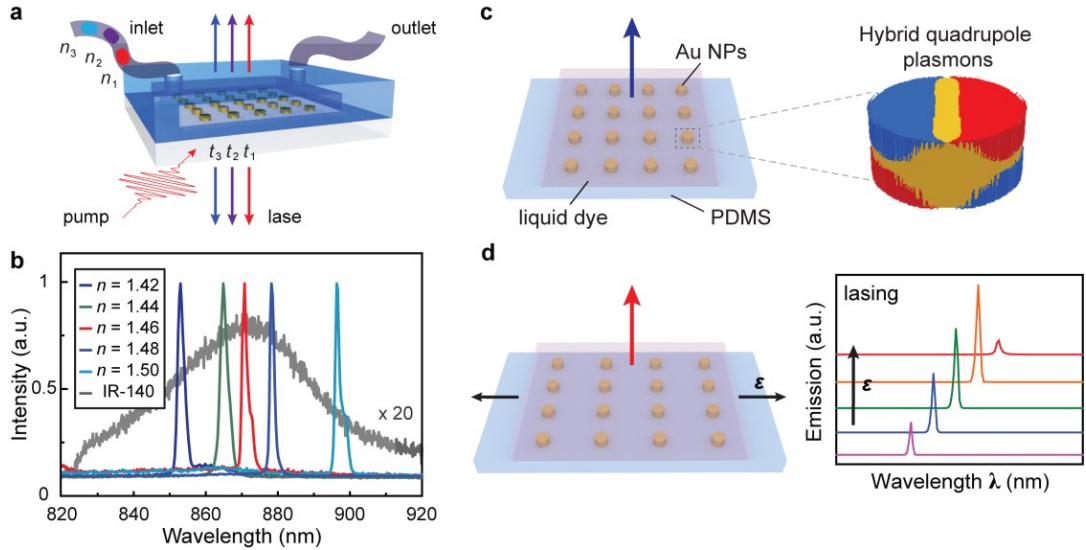

**Figure 6. Device tunability and optical multiplexing for nanolasers.** (**a**) Real-time tunable nanolaser with a microfluidic channel. (**b**) Measured dynamic change of lasing emission as a function of the refractive index of liquid dye solvent. (**c**) Scheme of a stretchable nanolaser with a Au NP lattice on an elastomeric substrate. (**d**) Wavelength-tunable lasing with strain $\varepsilon$ imposed on the substrate. Panels (a-b) adapted from ref. 53 with permission; copyright 2015 Nature Publishing. Panel (c-d) adapted from ref. 71 with permission; copyright 2018 American Chemical Society.

**3.3 Electrical pump**

For future applications of miniaturized lasers in optical communications, information processing, and on-chip circuits, electrical pump of the system is in high demand. Semiconductors as the gain media have been integrated into microscale optical cavities to be compatible with electrical pump and for better photostability and device efficiency. An electrically driven, single-mode laser at wavelength size was realized from photonic band gap modes at room temperature with a threshold current of 260 µA [9]. Lower lasing threshold was later demonstrated in a photonic-crystal nanocavity laser integrated with InAs quantum dot, pumped by a lateral p–i–n junction formed by ion implantation (**Figure 7a**). CW lasing was observed at temperatures up to 150 K with a threshold of 287 nA [72]. Highly stable, electrically pumped laser in the ultraviolet regime was realized at low temperatures by AlGaN core–shell nanowire arrays monolithically grown on a Si substrate [73] (**Figure 7b**). The lasing threshold in tens of A·cm$^{-2}$ is a few orders of magnitude lower than those of previously reported quantum-well lasers based on mm-scale cavities [74]. Both materials development and structural designs are needed for the electrical pump of next-generation nanoscale lasers with an even lower lasing threshold.

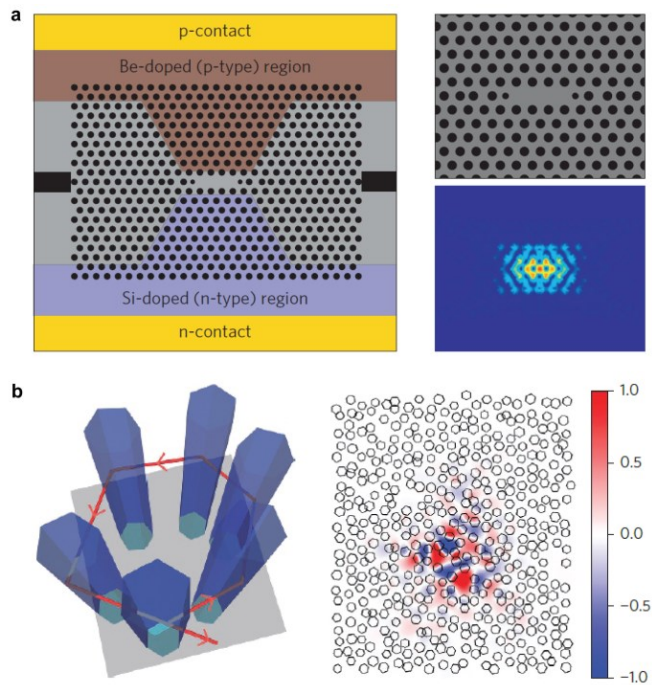

**Figure 7. Electrical pumping for small-scale optical cavities.** (**a**) Electrically pumped nanocavity laser based on quantum dot coupled to photonic crystal with ultralow lasing threshold. (**b**) Electrically injected AlGaN nanowire ultraviolet lasers at low temperature with ultralow threshold. Panel (a) adapted from ref. 72 with permission; copyright 2011 Nature Publishing. Panel (b) adapted from ref. 73 with permission; copyright 2015 Nature Publishing.

## 4. EMERGING APPLICATIONS IN QUANTUM PHOTONICS

### 4.1 Quantum emitters coupled to photonic nanocavities

Quantum engineering is an emerging area that drives many communication, simulation, and computing applications. Manipulating light-matter interactions with structured nanomaterials offers opportunities to enhance quantum coherence by a large optical density of states at the optical resonances [75]. Quantum emitters in 2D materials and color center defects in diamond, Si and SiC are intriguing candidates for integrated on-chip quantum nanophotonics [76, 77]. For example, isolated defects embedded in 2D hexagonal boron nitride (hBN) can support single photon emission at room temperature [78]. However, realizing integrated nanophotonic systems requires coupling quantum emitters to well-defined optical cavities [25, 79].

A hybrid quantum system was developed in which quantum emitters in hBN are deterministically coupled to high-quality plasmonic nanocavity arrays with a wet transfer method (**Figure 8a**). **Figure 8b** shows a confocal photoluminescence (PL) map of a pristine hBN flake, where a selected single emitter was marked by a red circle [80]. The pre-characterized emitter was later transferred onto a plasmonic lattice with a matching SLR wavelength to obtain spectral overlap of emitter emission and cavity resonance. Emission of the emitter was enhanced by three folds with maintained single-photon character, as evident from the second-order coherence function $g^2(\tau)$ ($g^2(0) = 0.1$) (**Figures 8c-d**). The measured PL enhancement could be even higher according to the modeled Purcell enhancement by the plasmonic lattice, if both spatial and spectral overlap between the emitter and the nanocavity can be ensured.

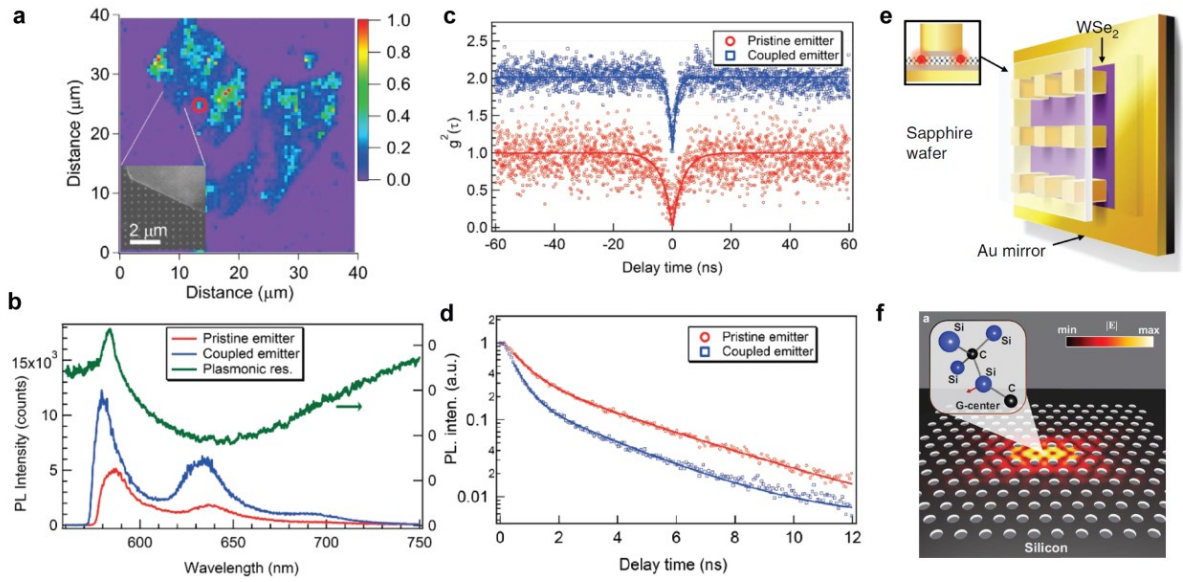

**Figure 8. Emerging applications of small-scale optical cavities in quantum photonics.** Various systems for deterministic coupling of quantum emitters to nanophotonic cavities. (**a**) PL confocal map of a hBN flake containing a single photon emitter (red circle). Inset representing SEM image of the hBN flake on a Au NP lattice. (**b**) PL spectra of the pristine and coupled single photon emitter. (**c**) Single-photon characteristics were maintained based on second-order correlation function $g^2(\tau)$. (**d**) Ultrafast dynamics measurements from the pristine and coupled systems. (**e**) Site-controlled quantum emitters in monolayer $WSe_2$ coupled to plasmonic nanogap modes. (**f**) Single atomic emissive center embedded in a silicon nanophotonic cavity. Panels (a-d) adapted from ref. 80 with permission; copyright 2017 American Chemical Society. Panel (e) adapted from ref. 81 with permission; copyright 2018 Nature Publishing. Panel (f) adapted from ref. 82 with permission; copyright 2023 Nature Publishing.

Single-photon enhancement can be further enhanced at lithographically defined locations based on the sharp corners of a metal nanocube, which deformed a 2D material and localized the emitters to electromagnetic hot spots (**Figure 8e**). Emitters were created onsite at plasmonic gap modes by the mechanically induced local strain [81]. The system sustains a Purcell factor up to 551, single-photon emission rates of up to 42 MHz, and a narrow exciton linewidth as low as 55 µeV. In addition, flux-grown $WSe_2$ increases the cavity-enhanced quantum yields up to 65%. **Figure 8f** shows a recent work that embedded atomic emissive center in a Si-based photonic crystal cavity [82]. Deterministic positioning and alignment of an atomic-scale defect and cavity dipole moments were enabled by statistically quantifying the dipole orientation of created atomic emissive center. The hybrid system showed 30-fold luminescence enhancement and an 8-fold acceleration of the emission rate.

**4.2 Future prospect**

Access to nanoscale and picoscale optical cavities opens many exciting possibilities in extreme light-matter interactions, atomic-cavity physics, and cavity photochemistry. Small-scale optical cavities can enhance quantum emission through the large optical density of states, and the ability to precisely position quantum emitters to the nanocavity hot spots will open prospects in hybrid quantum photonic circuits. Besides single emitters, a large-scale quantum condensation behavior among collective emitters was observed for plasmonic NP lattices coupled to dye molecule solutions [83]. Spectral and spatial mapping of luminescence suggests a transition from thermalization to lasing to Bose–Einstein condensation by tailoring the optical band structure. Such macroscopic quantum coherence suggests the unprecedented potential of nanocavity arrays because of their ultrafast, room-temperature, and on-chip nature.

Theoretical advances are driving the development of new optical cavities beyond conventional architectures. For example, a bound state in the continuum supports extremely high mode quality factors, and lasing action was demonstrated from an optically pumped bound states in the continuum cavity at room temperature [84, 85]. By exploiting topological states in a photonic crystal nanocavity array, lasing from such a system is geometry invariant and robust to local structural defects [86-88]. Such rationally structured cavities systems go beyond the conventional figure-of-merit matrix of laser devices and offer intriguing prospects in encoded information transportation, exceptional exciton-polariton behaviors, and on-chip robust photonic devices.

**AUTHOR INFORMATION**

**Corresponding Author**

*Phone: +1 (224) 420-6869. E-mail: danqingwang2018@u.northwestern.edu

**ORCID**

Danqing Wang: 0000-0002-7369-1944

Ankun Yang: 0000-0002-0274-4025

**ACKNOWLEDGMENTS**

This work was supported by the Miller research fellowship at University of California, Berkeley (D.W.).